%% file: rotation.tex
\documentclass[letterpaper,12pt,oneside]{amsart}
\usepackage[breaklinks=true,colorlinks=true,dvips=true]{hyperref}
\usepackage{amsmath,amssymb}
\usepackage{mathdef}
\usepackage{graphicx}

\input{style}

\begin{document}
\input{abstract}
\maketitle

\input{intro}
\input{gen}
\input{rot}
\input{concl}

\bibliographystyle{abbrv}
\bibliography{mech_en,mech_ru_en,calc_en}

\end{document}

%% file: style.tex
\title{On Lagrange Theory of Shells of Revolution}
\author{Vladimir V. Eliseev, Alexey S. Skoblikov}
\address{195426, Russia, Saint-Petersburg, pr.Kosygina, d.7, korp.1, kv.69}
\email{almays@almays.biz}

\keywords{elastic, shells of revolution, cylindrical, numerical comparison}

\date{}

\renewcommand{\L}{\mathcal{L}}
\newcommand{\D}{\mathcal{D}}
\newcommand{\df}{\partial_\phi}
\newcommand{\ds}{\partial_s}
\newcommand{\dz}{\partial_\zeta}
\newcommand{\kf}{k_\phi}
\newcommand{\kz}{k_z}

\newcommand{\ta}{\T a}
\newcommand{\tb}{\T b}
\newcommand{\tep}{\T \epsilon}
\newcommand{\tka}{\T \varkappa}
\newcommand{\tta}{\T \tau}
\newcommand{\tmu}{\T \mu}

\newcommand{\intO}[1]{\int_O #1\,dO}
\newcommand{\intdO}[1]{\oint_{\partial O} #1\,dl}
\newcommand{\into}[1]{\int #1\,dO}
\newcommand{\intdo}[1]{\oint #1\,dl}

%% file: abstract.tex
\begin{abstract}

The new linear theory of elastic shells is presented in this paper. This theory is free from various logical imperfections, that may be found in the approaches of earlier researchers. On the base of this theory the equations of shells of revolution are built. The equations of cylindrical shell are compared by a number of properties with the equations of other researchers and with elasticity equations of 3-dimensional tube.

Important quality decisions are stated.

\end{abstract}

%% file: intro.tex
\section{Introduction}

Tensor analysis in the theory of elasticity was widely used by A.~I.~Lurye
(for example, \cite{Lurye_1970_en}). The direct tensor analysis prevails: tensors
are considered as invariant objects, component form is used rarely, only when
necessary. This approach makes analytical calculations more comprehensible,
letting us pay more attention to the sense of formulas, not to algebra.

Recently great success was made in the theory of elastic shells. First approaches were made by Berdichevsky \cite{Berdichevsky_1983_en}, then direct tensor algebra was involved by V.\,V.\,Eliseev \cite[234]{Eliseev_1999_en}. Now it seems to be logically and mathematically perfect. It was not published in English yet, therefore we will first present this new approach, based on direct tensor analysis and principle of possible dislocations (Lagrange principle).

Having general theory of elastic shells we can build simplified forms of the theory. Here we derive the equations of shells of revolution. The equations of cylindrical shell are compared by a number of properties with the equations, derived by other researchers (\cite{Biderman_1977_en}, \cite{Vibrations_1978_en}, \cite{Vlasov_1951_naca}, \cite{Goldenveizer_1976_en}, \cite{Flugge_1962_en}) and with 3-dimensional model of the tube.

Important quality decisions are stated.

%% file: gen.tex
\section{General Theory}

The bases of the shell theory are considered to be well known \cite{Goldenveizer_1976_en}, \cite{Novozhilov_1962_en}, but its forming is not finished yet. Traditional approaches yield to asymptotic decomposition of 3D problem \cite{Goldenveizer_1976_en}, \cite{Berdichevsky_1983_en}; nevertheless the presentation of the shell as material surface doesn't lose its significance \cite{Zhilin_88_en}, \cite{Pietraszkiewicz_89_en}, \cite{Zubov_96_en}. In this study we consider linearly elastic shells as surfaces of particles with five degrees of freedom --- material normals. Rooting in methods of analytical mechanics and representation of stresses as Lagrange multipliers \cite{Rabotnov_1988_en}, it's easy to derive all equations of Kirchgoff-Love type theory. The results slightly differ from the already known.

\subsection{Geometry of material surface}

Radius-vector is the function of two coordinates $\V r(q^\alpha)$. Let's introduce vector basis in the tangent plane $\V r_\alpha \equiv \frac{\partial\V r}{\partial q^\alpha} \equiv \pd{\V r}{\alpha}$, normal ort is $\V n = \frac{\V r_1 \VE \V r_2}{|\V r_1 \VE \V r_2|}$, reciprocal basis is $\V r^\alpha$ ($\V r^\alpha \SC \V r_\beta=\delta_\beta^\alpha$), Hamilton operator $\nabla=\V r^\alpha \pd{}{\alpha}$, and metric tensors are $\T a=\nabla\V r = \V r^\alpha\V r_\alpha$ and $\T b = -\nabla\V n = -\V r^\alpha\V n_\alpha$, $\V n_\alpha \equiv \pd{\V n}{\alpha}$. Derivative formulas
\begin{equation*}
\begin{aligned}
\pd{\V r_\beta}{\alpha} &\equiv \V r_{\alpha\beta} = \Gamma_{\alpha\beta\lambda}\V r^\lambda + b_{\alpha\beta}\V n,\qquad
	\V n_\alpha = -b_{\alpha\beta}\V r^\beta\\
	\Gamma_{\alpha\beta\lambda} &= \frac{1}{2} \left(\pd{a_{\beta\lambda}}{\alpha} +
										\pd{a_{\alpha\lambda}}{\beta} - \pd{a_{\alpha\beta}}{\lambda}\right)
\end{aligned}
\end{equation*}
show that the form of the surface is totally defined by the dependency of covariant components $a_{\alpha\beta}$ and $b_{\alpha\beta}$ from coordinates \cite{Goldenveizer_1976_en}.

It's important to define precisely the degrees of freedom of the particles of material surface. The particles of momentless shell have only three degrees --- translation. In the Cosserat model particles are solid bodies with six degrees. But classical views of Kirhgoff-Love imply another model: particles are material normals with five degrees. Vector of small rotation $\V \theta$ in this case lies in the tangent plane, like all moments. Small change of normal during deformation and work are

\begin{equation*}
\dot{\V n} = \V \theta \VE \V n \equiv \V \phi, \quad
	\V m \SC \V \theta = \V m' \SC \V \phi, \quad \V m' \equiv \V m \VE \V n
\end{equation*}
(it's more convenient to work with $\V \phi$ rather than with $\V \theta$).

Vector of displacement $\V u$ --- is the increment of radius-vector ($\dot{\V r}$). Its connection with rotation follows from orthogonality
\begin{equation}
\label{eqConstraint}
\V r_\alpha \SC \V n = 0 \quad \Rightarrow \quad
	\V u_\alpha \SC \V n + \V r_\alpha \SC \V\phi = 0 \quad \Rightarrow \quad
	\V\phi = -\grad \V u \SC \V n.
\end{equation}

The form of the considered material surface is defined by covariant components $a_{\alpha\beta}$ and $b_{\alpha\beta}$, its small deformations are characterized by tensors
\begin{equation*}
\tep \equiv \frac{1}{2} \dot a_{\alpha\beta} \V r^\alpha \V r^\beta, \qquad
	\tka \equiv \dot b_{\alpha\beta} \V r^\alpha\V r^\beta.
\end{equation*}

Evaluating the components' increments we get
\begin{equation}
\label{eqDeform}
	\begin{aligned}
		a_{\alpha\beta} =& \V r_\alpha \SC \V r_\beta \quad \Rightarrow \quad
			\dot a_{\alpha\beta}=\V u_\alpha \SC \V r_\beta + \V r_\alpha \SC \V u_\beta \quad \Rightarrow \\
		&\Rightarrow \quad \tep = \frac{1}{2} \left(\grad \V u \SC \ta + \ta \SC \Trans{\grad\V u}\right)
			\equiv(\Sym{\grad\V u})_\| \\
		b_{\alpha\beta} =& -\V n_\alpha \SC \V r_\beta \quad \Rightarrow \quad
			\dot b_{\alpha\beta} = -\V\phi_\alpha \SC \V r_\beta - \V n_\alpha \SC \V u_\beta \quad \Rightarrow \\
		&\Rightarrow \quad \tka = -\left(\grad\V\phi\right)_\| + \tb \SC \Trans{\grad\V u}
	\end{aligned}
\end{equation}
(sign $(\ldots)_\|$ means belonging to the tangent plane or can also be expressed as $\V v \SC \ta$ for vectors and $\ta\SC\T T\SC\ta$ for tensors of rank 2).

Strain tensor $\tep$, defining the change of lengths and angles on the surface, takes place in all presentations of the shell theory, but tensor $\tka$ in the form \eqref{eqDeform} is not so well-known. Let's note that both tensors are symmetric and belong to the tangent plane. Due to the relation \eqref{eqConstraint} tensor of bending and rotation $\tka$ can be presented like:
\begin{equation*}
	\tka = \left(\grad\grad\V u \SC \V n\right)_\|
\end{equation*}
(it resembles classic theory of plates).

\subsection{Virtual work principle and its consequences}

It's possible to derive all system of equations of elastic system from this differential variational principle (except equations caused by material structure). For the considered model of the shell the following form of the principle is obvious:
\begin{equation}
\label{eqVWP}
	\intO{\left(\V f \SC \delta\V u + \V m' \SC \delta\V\phi - \delta\Pi\right)} +
		\intdO{\left(\V F \SC \delta\V u + \V M \SC \delta\V\phi\right)} = 0.
\end{equation}
Here $\V f$ and $\V m'$ --- force and moment loads per unit area (in dynamics they contain inertia terms), $\V F$ and $\V M$ --- loads on the boundary contour $\partial O$, $\Pi$ --- the density of energy of deformation.

For getting the stress tensors, deriving the equilibrium equations and Cauchy type formulas, let's use the technique, suggested in \cite{Rabotnov_1988_en}. On virtual dislocations without deformation
\begin{equation}
\label{eqConstrRigid}
	\delta\Pi = 0,
		\qquad \underline{\delta\tep = 0}, \qquad \underline{\delta\tka = 0}.
\end{equation}

After taking into account these equations in \eqref{eqVWP} we get variational problem with constraints (underlined). Introducing corresponding Lagrange multipliers the following equation arises
\begin{equation}
\label{eqVWP2}
	\begin{split}
		\into{\left[\V f \SC \delta\V u + \V m' \SC \delta\V\phi -
			\tta \SC \delta\tep - \tmu \SC \delta\tka -
			\V\lambda \SC \left(\delta\V\phi + \grad\delta\V u \SC \V n\right)\right]}\\
		+\intdo{\left(\V F\SC\delta\V u + \V M^0 \SC \delta\V\phi\right)} = 0.
	\end{split}
\end{equation}
Tensors $\tta$ and $\tmu$ are symmetric and belong to the tangent plane
--- such are constraints \eqref{eqConstrRigid}. Vector multiplier $\V\lambda\ (=\V\lambda_\|)$ is introduced according to \eqref{eqConstraint} and allows to vary independently $\V u$ and $\V\phi$.

Using the identities
\begin{equation*}
	\begin{aligned}
		\tta \SC \tep &= \div\left(\tta \SC \V u\right) -
			\div \tta \SC\V u\\
		\tmu \SC \tka &= \div\left(-\tmu \SC \V\phi +
			\tmu \SC \tb \SC \V u\right) + \div\tmu \SC \V\phi -
			\div\left(\tmu \SC \tb\right) \SC \V u\\
		\V\lambda \SC \grad\V u \SC \V n &= \div\left(\V\lambda\V n \SC \V u\right) -
			\div\left(\V\lambda\V n\right) \SC \V u
	\end{aligned}
\end{equation*}
and theorem of divergence
\begin{equation*}
	\intdO{\V\nu \SC \V u} = \intO{\left(\div \V u + 2H \V n \SC \V u\right)}
\end{equation*}
($\V\nu$ --- ort, normal to $\partial O$ in the tangent plane,
$2H=\tr\tb$ --- average curvature, $\V u$ may be the tensor of any rank),
let's transform \eqref{eqVWP2}:
\begin{equation}
\label{eqBalance}
	\begin{aligned}
		\int \biggl[&\left\{\V f + \div \left(\tta + \tmu \SC \tb +
			\V\lambda\V n\right)\right\}\SC\delta\V u\\
		&+ \left\{\V m' - \left(\div\tmu\right)_\| - \V\lambda\right\} \SC \delta\V\phi\biggr]\,dO \\
		+ \oint\biggl[&\left\{\V F - \V\nu \SC \left(\tta + \tmu \SC \tb + \V\lambda\V n\right)\right\}
			\SC\delta\V u\\
		&+ \left\{\V M + \V\nu \SC \tmu\right\}\SC\delta\V\phi\biggr]\,dl = 0.
	\end{aligned}
\end{equation}

Variations of $\delta\V u$ and $\delta\V\phi$ in the area and on the boundary are arbitrary, therefore the vectors in the curly brackets are zeros --- that is the equilibrium equations and Cauchy type formulas (for $\V F$ and $\V M$). This derivation relates not only to elastic behavior, term $(-\delta\Pi)$ just denominates the virtual work of internal forces per unit area.

But for elastic shell principle \eqref{eqVWP} allows to get the elasticity relations. Taking into account in \eqref{eqVWP} the consequences \eqref{eqBalance}, we get:
\begin{gather*}
		\delta\Pi = \tta \SC \delta\tep + \tmu \SC \delta\tka \quad\Rightarrow \\
		\Rightarrow \quad \Pi = \Pi(\tep, \tka),\quad
			\tta = \frac{\partial\Pi}{\partial\tep}, \quad
			\tmu = \frac{\partial\Pi}{\partial\tka}.
\end{gather*}
Elastic potential $\Pi$ in the linear theory is quadratic form, the number of coefficients in the case of general anisotropy is 21. In the simplest case of isotropy without cross couplings we get the relations with four constants
\begin{equation}
\label{eqHookSimple}
	\begin{gathered}
		2\Pi = C_1\epsilon^2 + C_2\tep \SC \tep +
			D_1\varkappa^2 + D_2\tka \SC \tka \\
		\epsilon \equiv \tr\tep, \qquad \varkappa \equiv \tr\tka\\
		\tta = C_1\epsilon\ta + C_2\tep, \quad	\tmu = D_1\varkappa\ta + D_2\tka.
	\end{gathered}
\end{equation}

In the limits of the above-said it is impossible to find the moduli $C_1$--$D_2$ in elastic relations \eqref{eqHookSimple} --- we should refer to asymptotic analysis of 3D problem by small thickness. But in the case of homogeneous isotropic material and independence of moduli from curvature we can take them like in the plate under plane stress and bending.
\begin{align*}
C_1 &= \frac{2\mu\nu h}{1-\nu} & C_2 &= 2\mu h \\
D_1 &= \frac{h^2}{12}C_1 & D_2 &= \frac{h^2}{12}C_2.
\end{align*}

\subsection{System of equations and boundary conditions}

Elasticity relations
\begin{equation}
	\label{eqBalanceStd}
	\left\{
	\begin{aligned}
		&\div\left(\tta + \tmu \SC \tb + \V Q\V n\right) + \V f = 0 \\
		&\left(\div\tmu\right)_\| + \V Q - \V m' = 0
	\end{aligned}
	\right.
\end{equation}
(where $\V Q \equiv \V\lambda$ --- shear forces) and \eqref{eqHookSimple} resemble the known ones (about which, however, in the literature there is no common opinion), and coincide only with \cite{Berdichevsky_1983_en}.

Boundary conditions, presented by the contour integral
\begin{equation}
	\label{eqBC}
	\intdo{\left[\left(\V F - \V\nu\SC\left(\tta + \tmu \SC \tb + \V Q\V n\right)\right)\SC\delta\V u +
		\left(\V M^0 + \V\nu\SC\tmu\right)\SC\delta\V\phi\right]} = 0,
\end{equation}
require reforming. Let's introduce the tangent ort on contour $\partial O$ --- $\V l = \V n \VE \V\nu$. After partial integration we have
\begin{gather*}
		\grad\bigl|_{\partial O} = \V\nu\pd{}{\nu}+\V l \pd{}{l},\qquad
			\V\phi\bigl|_{\partial O} = -\left(\V\nu\pd{\V u}{\nu} + \V l\pd{\V u}{l}\right)\SC\V n\\
		\intdo{\V A\SC\V\phi} =
			\intdo{\left[-\V A\SC\V\nu\V n\SC\pd{\V u}{\nu} +
				\pd{\left(\V A\SC\V l\V n\right)}{l}\SC\V u\right]}.
\end{gather*}
Transforming in this way the integral \eqref{eqBC}, we get the boundary conditions in the general form
\begin{equation}
		\left[\V F - \V\nu\SC\left(\tta + \tmu \SC \tb + \V Q\V n\right) + \pd{\left(\V A\SC\V l\V n\right)}{l}\right]
		\SC\delta\V u - \V A\SC\V\nu\V n\SC\pd{\delta\V u}{\nu} = 0
\end{equation}
where $\V A\equiv\V M + \V\nu\SC\tmu$.

On the free edge the vector $\delta\V u$ and scalar $\V n\SC\pd{\delta\V u}{\nu}$ are unrestricted --- their multipliers must be zeros. On the restrained edge $\V u$ and $\V n\SC\pd{\V u}{\nu}$ are defined --- in all cases, including mixed ones, we have four conditions in components.

It should be noted that during partial integration the contour was supposed to be smooth. Bit if there are corner points, then value $\pd{\left(\V A\SC\V l\V n\right)}{l}$ contains $\delta$-functions --- like with point force in the load $\V F$. This well-known fact was being explained by many authors.

\subsection{Consistency equations and statical-geometrical analogy}

Components $a_{\alpha\beta}$ and $b_{\alpha\beta}$ must satisfy three differential equations of Gauss-Peterson-Kodacci \cite{Goldenveizer_1976_en} (following from symmetry $\V r_{\alpha\beta\gamma}$ by any pair of indexes). Their variation leads to the consistency equations for $\tep$ and $\tka$. Let's consider another derivation, connected with unambiguity of displacements and rotations. Let's base on the theorem of circulation:
\begin{equation}
\label{eqCircularThConseq}
\begin{gathered}
	\intdO{\left(d\V r\SC\T v\right)} = \intO{\V n\SC\rot\T v}\\
	\T v = \grad\V u: \quad \V n\SC\rot\T v = 0 \quad
		\Leftrightarrow \quad \div\left(\V n \VE \T v\right) = 0.
\end{gathered}
\end{equation}

Let's take the gradient of displacement in the form
\begin{equation}
	\label{eqGradU}
	\grad\V u = \tep - \ta\VE\V\Omega, \qquad
		\V\Omega\equiv\rot\V u\SC\left(\ta + \frac{1}{2}\V n\V n\right) =
		\Omega_n\V n + \V n \VE \V\phi.
\end{equation}
By the reason of $d\V u = \tep\SC d\V r + \V\Omega\VE d\V r$,
$\V\Omega$ is the vector of small rotation; but let's note that generalized coordinates has only its ``plain'' part
$\V\Omega_\| = \V n\VE\V\phi = \V\theta$. Substitution of $\grad\V u$
from \eqref{eqGradU} in \eqref{eqCircularThConseq} leads to
\begin{equation}
\label{eqNablaOmegaN}
	\begin{gathered}
	\div\left(\V n\VE\tep\right) - \V n\div\V\Omega	+ \grad\V\Omega\SC\V n = 0 \quad\Rightarrow \\
	\Rightarrow \quad \grad\V\Omega\SC\V n = -(\div(\V n\VE\tep))_\|.
	\end{gathered}
\end{equation}

Next let's consider ``moment'' tensor of deformation
\begin{equation*}
	\tka = (\grad\grad\V u\SC\V n)_\| = \tb\SC\tep - \grad\V\Omega\VE\V n.
\end{equation*}
Together with \eqref{eqNablaOmegaN} it yields to
\begin{equation*}
	\grad\V\Omega = (\tka - \tb\SC\tep)\VE\V n - (\div(\V n\VE\tep))_\|\V n.
\end{equation*}
The condition \eqref{eqCircularThConseq} for $\grad\V\Omega$ can be transformed to
\begin{equation}
	\label{eqCompliance}
	\div(\tka^* - \tb^*\SC\tep^* + \V\Lambda\V n) = 0, \qquad (\div\tep^*)_\| - \V\Lambda = 0,
\end{equation}
where $(\ldots)^*$ means the rotation of the tensor: $\tka^* = -\V n\VE\tka\VE\V n$ etc.

Equations \eqref{eqCompliance} express the consistency of deformation. $\V\Lambda$ --- is just a denotation, therefore we have three equations in components. Comparing \eqref{eqCompliance} with \eqref{eqBalanceStd}, we see total analogy between equilibrium and consistency equations. Statical-geometrical analogy is known already, but not in this form.

%% file: rot.tex
\section{Shells of revolution}

\subsection{General equations}

Let's define the coordinate system on the shell: $\{\phi,s\}$, $\phi$ is the angular
coordinate of meridional plane (longitude), $s$ --- arc coordinate along the meridian.

Let's express radius-vector in the form
\begin{equation*}
  \V r = x(s)\V i + R(s)\V\rho(\phi).
\end{equation*}
It should be noted that $\pd{\V\rho}{\phi}=\V e_\phi,\ \pd{\V
e_\phi}{\phi}=-\V\rho$. Reference vectors in this coordinate system are:
\begin{equation*}
  \begin{aligned}
    \V r_\phi &= R(s)\V e_\phi \\
    \V r_s    &= x'(s)\V i + R'(s)\V\rho.
  \end{aligned}
\end{equation*}

Let's pay attention to the following point: $x$ and $R$ are cartesian coordinates of meridian in the meridional plane, therefore $ds^2=dx^2+dR^2$.
Dividing both parts of this expression by $ds^2$, we get
\begin{equation*}
  x'(s)^2 + R'(s)^2 = 1.
\end{equation*}
The values $x'(s)$ and $R'(s)$ satisfy the trigonometric identity, that's why the following denotation can be introduced: $R'(s)=\sin\psi(s),\ x'(s)=\cos\psi(s)$.
The geometrical sense of $\psi(s)$ is the angle between tangent to meridian $\V r_s$ and symmetry axis $\V i$. Now reference vectors and normal may be written as
\begin{equation*}
  \begin{aligned}
    \V r_\phi &= R\V e_\phi \\
    \V r_s    &= \V i \cos\psi + \V\rho \sin\psi \equiv \V t \\
    \V n      &= {\V e}_\phi\VE\V t = \V\rho \cos\psi - \V i \sin\psi.
  \end{aligned}
\end{equation*}
Derivation formulas are
\begin{equation*}
  \begin{aligned}
    \pd{\V e_\phi}{\phi} &= -\V\rho = -\V t\sin\psi-\V n\cos\psi &\qquad \pd{\V e_\phi}{s} &= 0 \\
    \pd{\V t}{\phi} &= \cos\psi\V e_\phi & \pd{\V t}{s} &= \psi'\V n \\
    \pd{\V n}{\phi} &= \sin\psi\V e_\phi & \pd{\V n}{s} &= -\psi'\V t.
  \end{aligned}
\end{equation*}
Introducing denotation $R^{-1}=k$, let's build reciprocal basis and Hamilton operator
\begin{equation*}
  \begin{aligned}
    \V r^\phi &= k\V e_\phi \\
    \V r^s    &= \V t \\
    \nabla &= k\V e_\phi \pd{}{\phi} + \V t\pd{}{s}.
  \end{aligned}
\end{equation*}
Finally let's calculate metric tensors of the shell of revolution
\begin{equation*}
  \begin{aligned}
    \T a &= \grad\V r = \T I - \V n\V n = \V e_\phi\V e_\phi + \V t\V t \\
    \T b &= -\grad\V n = -k\sin\psi\V e_\phi\V e_\phi + \psi'\V t\V t.
  \end{aligned}
\end{equation*}

The equilibrium equations can be derived from the general ones easily. We'll omit all further calculations. Here are only formulas for the gradient of vector and divergence of tensor.
\begin{equation*}
  \begin{aligned}
    \grad\V u &= \V e_\phi\V e_\phi \left( k\left( \pd{u_\phi}{\phi}
+ u_s \sin\psi + u_n \cos\psi \right) \right) + \V t\V t \left( \pd{u_s}{s} -
\psi' u_n \right) \\&{}
    + \V e_\phi\V t \left( k \left( \pd{u_s}{\phi}
- u_\phi \sin\psi \right) \right) + \V t\V e_\phi \left( \pd{u_\phi}{s}
\right) \\&{}
    + \V e_\phi\V n \left( k \left( \pd{u_n}{\phi}
- u_\phi \cos\psi \right) \right) + \V t\V n \left( \pd{u_n}{s}
+ u_s \psi' \right) \\
    \div\T\mu &= \V e_\phi \left( k \left( \pd{\mu_\phi}{\phi}
+ \mu_{\phi s} \sin\psi + \mu_{s\phi} \sin\psi \right) + \pd{\mu_{s\phi}}{s}
\right) \\&{}
    + \V t \left( k \left( -\mu_\phi \sin\psi + \mu_s \sin\psi
+ \pd{\mu_{\phi s}}{\phi} \right) +\pd{\mu_s}{s} \right) \\&{}
    + \V n \left( -k \mu_\phi \cos\psi + \mu_s \psi' \right).
  \end{aligned}
\end{equation*}

On the base of these expressions and derivation formulas we can build the following sequence of precise equations:
\begin{subequations}\label{eq:rot}
\begin{gather}
\intertext{kinematic relations}
  \begin{aligned}\label{eq:rot_deform}
    \epsilon_\phi &= k \left( \pd{u_\phi}{\phi} + u_s \sin\psi
+ u_n \cos\psi \right) \\
    \epsilon_s &= \pd{u_s}{s} - u_n \psi' \\
    \epsilon_{\phi s}&=\epsilon_{s\phi}=\frac{1}{2}\left( k\left( \pd{u_s}{\phi}
- u_\phi \sin\psi \right) +\pd{u_\phi}{s} \right)
  \end{aligned} \\
  \begin{aligned}
    \varkappa_\phi &= k^2 \left( \pd[2]{u_n}{\phi\phi}
- 2\pd{u_\phi}{\phi} \cos\psi - u_s \sin\psi\cos\psi - u_n \cos^2\psi \right) \\&{}
+ k \left( \pd{u_n}{s} \sin\psi +u_s \psi'\sin\psi \right) \\
    \varkappa_s &= \pd[2]{u_n}{ss} + 2\pd{u_s}{s} \psi' + u_s \psi'' - u_n \psi'' \\
    \varkappa_{\phi s} &= k^2 \left( -\pd{u_n}{\phi} \sin\psi
+ u_\phi \sin\psi\cos\psi \right) \\&{}
+ k \left( \pd[2]{u_n}{s\phi} +
\pd{u_s}{\phi} \psi'
  - \pd{u_\phi}{s} \cos\psi \right) \\
    \varkappa_{s\phi} &= k \left( \pd[2]{u_n}{\phi s} - \pd{u_\phi}{s} cos\psi
+ \pd{u_s}{\phi} \psi' \right) + k' \left( \pd{u_n}{\phi} - u_\phi \cos\psi
\right)
  \end{aligned} \\
\intertext{elasticity relations}\label{eq:rot_hook}
  \tau_\phi = C_1 \left( \epsilon_\phi+\epsilon_s \right)+C_2\epsilon_\phi \qquad
  \tau_s = C_1 \left( \epsilon_\phi+\epsilon_s \right)+C_2\epsilon_s \qquad
  \tau_{\phi s} = \tau_{s \phi} = C_2\epsilon_{\phi s} \\
\begin{aligned}
  \mu_\phi &= D_1 \left( \varkappa_\phi+\varkappa_s \right)+D_2\varkappa_\phi &
  \mu_s &= D_1 \left( \varkappa_\phi+\varkappa_s \right)+D_2\varkappa_s \\
  \mu_{\phi s} &= D_2\varkappa_{\phi s} &
  \mu_{s\phi} &= D_2\varkappa_{s\phi}
\end{aligned} \\
\intertext{shear forces}
  \begin{aligned}\label{eq:rot_cut}
    Q_\phi &= - k \left( \pd{\mu_\phi}{\phi} + \mu_{\phi s} \sin\psi
+ \mu_{s\phi} \sin\psi \right) - \pd{\mu_{s\phi}}{s} + m_s \\
    Q_s &= -k \left( -\mu_\phi \sin\psi + \mu_s \sin\psi
+ \pd{\mu_{\phi s}}{\phi} \right) - \pd{\mu_s}{s} - m_\phi
  \end{aligned} \\
\intertext{equilibrium equations}
  \left\{
  \begin{aligned}\label{eq:rot_equil}
    & k^2 \left( -\pd{\mu_\phi}{\phi} \cos\psi
- \mu_{s\phi} \cos\psi \sin\psi \right) \\&\quad{}
    + k \Bigl( \pd{\tau_\phi}{\phi}
+ 2 \tau_{\phi s} \sin\psi + \mu_{\phi s} \psi'\sin\psi - \pd{\mu_{s\phi}}{s}
\cos\psi \\&\qquad{}
    + \mu_{s\phi} \psi'\sin\psi + Q_\phi \cos\psi \Bigr)
+ \pd{\tau_{\phi s}}{s} - k'\mu_{s\phi} \cos\psi + f_\phi = 0 \\
    & k^2 \mu_\phi \cos\psi\sin\psi + k \left( -\tau_\phi \sin\psi
+ \tau_s \sin\psi + \pd{\tau_{\phi s}}{\phi} + \mu_s \psi'\sin\psi +
\pd{\mu_{\phi s}}{\phi} \psi' \right) \\&\quad{}
    + \pd{\tau_s}{s}
+ \pd{\mu_s}{s} \psi' + \mu_s \psi'' - Q_s \psi' + f_s = 0 \\
    & k^2 \mu_\phi \cos^2\psi + k \left( -\tau_\phi \cos\psi
+ \pd{Q_\phi}{\phi} + Q_s \sin\psi \right) \\&\quad{}
    + \tau_s \psi' + \mu_s \psi'{^2} + \pd{Q_s}{s} + f_n = 0.
  \end{aligned}
  \right.
\end{gather}
\end{subequations}

Let's note that tensors $\tka$ and $\tmu$ are symmetric like $\tta$ and $\tep$. It becomes obvious that $\varkappa_{\phi s}=\varkappa_{s\phi}$ if we return from notation $\psi(s),\ k(s)$ to $R(s),\ x(s)$. The equations for displacements can be derived from equations \eqref{eq:rot}. Stress equations contain \eqref{eq:rot_equil} and consistency equations \eqref{eqCompliance}, which in the case of shell of revolution are
\begin{subequations}\label{eq:rot_unbreak}
\begin{gather}
  \left\{
  \begin{aligned}
    & - k^2 \epsilon_{\phi s} \cos\psi\sin\psi \\&\quad{}
    + k \left( \pd{\varkappa_s}{\phi}
- \varkappa_{s\phi} \sin\psi - \varkappa_{\phi s} \sin\psi - \pd{\epsilon_s}{\phi}
\psi' + \pd{\epsilon_{\phi s}}{s} \cos\psi + \Lambda_\phi \cos\psi \right) \\&\quad{}
    + k' \epsilon_{\phi s} \cos\psi
- \pd{\varkappa_{\phi s}}{s} = 0 \\
    & k^2 \epsilon_\phi \cos\psi\sin\psi \\&\quad{}
    + k \Bigl( - \varkappa_s \sin\psi
+ \varkappa_\phi \sin\psi - \pd{\varkappa_{s\phi}}{\phi} + \epsilon_s \psi'\sin\psi
+ \pd{\epsilon_{\phi s}}{\phi} \\&\qquad{}
    + \pd{\epsilon_\phi}{s} \cos\psi
-\epsilon_\phi \psi'\sin\psi \Bigr)+k' \epsilon_\phi \cos\psi
+ \pd{\varkappa_\phi}{s} - \Lambda_s\psi' = 0 \\
    & k \left( - \varkappa_s \cos\psi - \epsilon_s \psi'\cos\psi
- \epsilon_\phi \psi'\cos\psi + \pd{\Lambda_\phi}{\phi} + \Lambda_s \sin\psi
\right) + \varkappa_\phi \psi' + \pd{\Lambda_s}{s} = 0,
  \end{aligned}
  \right.
\intertext{where}
  \begin{aligned}
    \Lambda_\phi &= k \left( \pd{\epsilon_s}{\phi}
- 2 \epsilon_{\phi s} \sin\psi \right) - \pd{\epsilon_{\phi s}}{s} \\
    \Lambda_s &= k \left( -\epsilon_s \sin\psi + \epsilon_\phi \sin\psi
- \pd{\epsilon_{\phi s}}{\phi}  \right) + \pd{\epsilon_\phi}{s},
  \end{aligned}
\intertext{and strains are expressed through stresses by formulas}
\begin{aligned}
  \varkappa_\phi &= A_1 \left( \mu_\phi+\mu_s \right)+A_2\mu_\phi &
  \varkappa_s &= A_1 \left( \mu_\phi+\mu_s \right)+A_2\mu_s \\
  \varkappa_{\phi s} &= A_2\mu_{\phi s} &
  \varkappa_{s\phi} &= A_2\mu_{s\phi}
\end{aligned} \\
  \epsilon_\phi = B_1 \left( \tau_\phi+\tau_s \right)+B_2\tau_\phi \qquad
  \epsilon_s = B_1 \left( \tau_\phi+\tau_s \right)+B_2\tau_s \qquad
  \epsilon_{\phi s} = B_2\tau_{\phi s}.
\end{gather}
\end{subequations}

Here $B_1=-\frac{C_1}{C_2\left( 2C_1+C_2 \right)},\ B_2=\frac{1}{C_2},\
A_1=-\frac{D_1}{D_2\left( 2D_1+D_2 \right)},\ A_2=\frac{1}{D_2}$.

Using formulas \eqref{eq:rot}, given in this section, we'll get the equations of cilyndrical shell, analyse them and compare by a number of properties with the equations of cilyndrical shells, given by other researchers (\cite{Biderman_1977_en}, \cite{Vibrations_1978_en}, \cite{Vlasov_1951_naca}, \cite{Goldenveizer_1976_en}, \cite{Flugge_1962_en}) and with 3-dimensional model of the tube.

\subsection{Cylindrical shell}\label{sec:rot_buil_cyl}

\subsubsection{Equations}

It's not difficult to derive the equations of cylindrical shell from the equations of shells of revolution \eqref{eq:rot}. That's enough to define the functions $x(s)$ and $R(s)$ appropriately.
\begin{equation*}
  x(s)=s \equiv z, \quad R(s)=R.
\end{equation*}

Function $\psi(s)$ can be calculated by formula
$\psi(s)=\arctan\frac{R'(s)}{x'(s)}=\arctan\frac{dR}{dx}$. In this case it's obvious, that $\psi=0$.

Cylindrical shell is famous for its equations for displacements are differential equations with \emph{constant coefficients}. Let's change the variable for expression simplification $s=\zeta R, \ds=\frac{1}{R}\dz$.
System \eqref{eq:rot} appears as
\begin{subequations}\label{eq:cyl}
\begin{gather}
  \L U(\phi,\zeta) + B(\phi,\zeta) = 0.
\intertext{Here}
\begin{aligned}
  U(\phi,\zeta)&=
\begin{pmatrix}
  u_\phi (\phi,\zeta) \\
  u_s (\phi,\zeta) \\
  u_n (\phi,\zeta)
\end{pmatrix}=
\begin{pmatrix}
  u_1 (\phi,\zeta) \\
  u_2 (\phi,\zeta) \\
  u_3(\phi,\zeta)
\end{pmatrix}\\
  B(\phi,\zeta)&=
\begin{pmatrix}
  b_\phi (\phi,\zeta) \\
  b_s (\phi,\zeta) \\
  b_n (\phi,\zeta)
\end{pmatrix}=
\begin{pmatrix}
  b_1 (\phi,\zeta) \\
  b_2 (\phi,\zeta) \\
  b_3 (\phi,\zeta)
\end{pmatrix}\\
  \L&=
\begin{pmatrix}
  L_{11} & L_{12} & L_{13} \\
  L_{21} & L_{22} & L_{23} \\
  L_{31} & L_{32} & L_{33}
\end{pmatrix},
\end{aligned} \\
\intertext{where}
  \begin{aligned}
    L_{11} &= \left(\frac{1}{2}+2\gamma\right)
\left( \dz^2 + 2\df^2 - \nu\dz^2 \right) &\quad
    L_{12} &= L_{21} = \frac{1}{2}(1+\nu)\dz\df \\
    L_{22} &= \frac{1}{2}\left( 2\dz^2 + \df^2 - \nu\df^2 \right) &
    L_{23} &= L_{32} = \nu\dz \\
    L_{33} &= 1+\gamma
\left( \lap\lap + 1 - 2\df^2-2\nu\dz^2 \right) &
    L_{31} &= L_{13} = \df + 2\gamma
\left( \df-\dz^2\df-\df^3 \right)
  \end{aligned} \\
  \begin{aligned}
    b_\phi(\phi,\zeta) &= \frac{R(1-\nu)}{2h\mu}
\left( R f_\phi(\phi,\zeta) + m_s(\phi,\zeta) \right) \\
    b_s(\phi,\zeta) &= \frac{R(1-\nu)}{2h\mu}R f_s(\phi,\zeta) \\
    b_n(\phi,\zeta) &= -\frac{R(1-\nu)}{2h\mu}
\left( R f_n(\phi,\zeta) - \pd{m_\phi(\phi,\zeta)}{\zeta} +
\pd{m_s(\phi,\zeta)}{\phi} \right).
  \end{aligned}
\end{gather}
\end{subequations}
Here $\nu,\ \mu$ --- correspondingly Poisson's ratio and shear module,
$\lap=\left(\df^2+\dz^2\right),\ \gamma=\frac{h^2}{12R^2}$.

\subsubsection{Roots of characteristic equation}

Let's consider sinusoidal loads and solutions\footnote{All the following is true also in the case of Fourier expansion of the solution $U(\phi,\zeta)=\sum_{m=0}^\infty
U^m(\zeta)\exp{mi\phi}$ with arbitrary loads.} $U(\phi,\zeta)=U^m(\zeta)\exp{mi\phi}$. After substituting such solution in the system \eqref{eq:cyl} and multiplying by $\exp{-mi\phi}$, we get the system of differential equations with constant coefficients for displacement intensities $U^m(\zeta)$. Trying complementary solution of the corresponding homogeneous system in the form $U^m(\zeta)=U^{mk}\exp{k\zeta}$, is equivalent to trying the solution of input homogeneous system \eqref{eq:cyl} in the form
\begin{equation}\label{eq:cyl_view}
  U(\phi,\zeta)=U^{mk}\exp{k\zeta+mi\phi}.
\end{equation}
After canceling by $\exp{k\zeta+mi\phi}$, we have the system of homogeneous algebraic equations for $\Trans{\left( U^{mk}_\phi,\ U^{mk}_\zeta,\ U^{mk}_n \right)}$. This system has solution only if its determinant equals to zero.

During trying the solution in the form \eqref{eq:cyl_view} differential operators turn into factors
\begin{equation*}
  \df=im, \quad \dz=k.
\end{equation*}
Thus resolvability requirement for such system can be written as
\begin{equation*}
  \D(im,k)=0, \qquad \text{where} \quad \D(\partial_\phi, \partial_\zeta) = \det\L.
\end{equation*}

Trying the solutions for $U^m(\zeta)$ separately for each harmonic by $\phi$, let's represent this equation as equation for $k$. Coefficient $m$ will be the parameter.
\begin{equation}\begin{split}\label{eq:cyl_char}
  \D &= k^8 \left( -\frac{1}{2}\gamma - 2\gamma^2 \right)
+ k^6 \left( m^2\left( 2\gamma-(\nu-5)\gamma^2 \right)
  + \nu\gamma+4\nu\gamma^2 \right) \\&{}
+ k^4 \biggl( m^4\left( -3\gamma+2(\nu-2)\gamma^2 \right)
  +m^2\left( 3\gamma+2(2-\nu+\nu^2)\gamma^2 \right)
  -2\gamma^2 \\&\quad{}
    +\frac{4\nu^2-5}{2}\gamma+\frac{1}{2}(\nu-1)(1+\nu) \biggr) \\&{}
+ k^2 \Bigl( m^6\left( 2\gamma-(\nu-1)\gamma^2 \right)
  + m^4\left( 2(\nu-1)\gamma^2-(4+\nu)\gamma \right) \\&\quad{}
    + m^2\left( -(\nu-1)\gamma^2+(2+\nu)\gamma \right) \Bigr)
- m^8\frac{1}{2}\gamma + m^6\gamma - m^4\frac{1}{2}\gamma = 0.
\end{split}\end{equation}

Let's examine two cases: axisymmetric solution ($m=0$) and antisymmetric\footnote{Names ``axisymmetric'' and ``antisymmetric'' are quite conventional. Under additional conditions $m=0$ would mean either real axial symmetry, or torsion, and $m=1$ should cause the bending of the cylinder like a beam.} solution ($m=1$). The solutions of the characteristic equation
\eqref{eq:cyl_char} for $k$ in these cases are:
\begin{gather*}
\intertext{$m=0$}
  k_{1,2,3,4}^{(0)}=0; \quad
k_{5,6,7,8}^{(0)}=\pm\sqrt{\nu \pm \sqrt{\gamma^{-1}+1}\sqrt{\nu^2-1}} \\
\intertext{$m=1$}
\begin{aligned}
  k_{1,2,3,4}^{(1)}&=0; \quad
  k_{5,6,7,8}^{(1)}=\pm\sqrt{\frac{1}{1+4\gamma}
\left( 2+5\gamma+\nu+3\nu\gamma \pm \sqrt{A} \right) } \\
  A&=-\gamma^{-1}-5-4\gamma+9\gamma^2+4\nu+ 22\nu\gamma+30\nu\gamma^2 \\&\quad{}
+\nu^2\gamma^{-1}+9\nu^2+26\nu^2\gamma+25\nu^2\gamma^2.
\end{aligned}
\end{gather*}

Let's look at the asymptotic form of these roots by $\gamma\to 0$ (the shell is thin).
\begin{equation*}
\begin{aligned}
  k^{(0)} &\to \pm \left( \gamma^{-1/4} \sqrt{\pm\sqrt{\nu^2-1}}
+ \gamma^{1/4} \frac{\nu}{2\sqrt{\pm\sqrt{\nu^2-1}}} + \dots \right) \\
  k^{(1)} &\to \pm \left( \gamma^{-1/4} \sqrt{\pm\sqrt{\nu^2-1}}
+ \gamma^{1/4} \frac{2+\nu}{2\sqrt{\pm\sqrt{\nu^2-1}}} + \dots \right).
\end{aligned}
\end{equation*}

\subsubsection{Solution}\label{sssec:solution}

Let's consider the case of sinusoidal load by both coordinates
\begin{equation*}
	b_m(\phi,z) = B_m \exp{i \kf \phi + i \kz z} \quad (m=1,2,3=\phi,z,n).
\end{equation*}
Let's consider also endlessly long cylinder, so that the border conditions on the ends may be neglected. The border conditions in the second coordinate $\phi$ are periodical conditions and are satisfied automatically if we try the solution of the system \eqref{eq:cyl} in the form
\begin{equation*}
\begin{aligned}
	u_\phi(\phi,z) &= U_\phi \exp{i \kf \phi + i \kz z} \\
	u_z(\phi,z) &= U_z \exp{i \kf \phi + i \kz z} \\
	u_n(\phi,z) &= \frac{U_n}{i} \exp{i \kf \phi + i \kz z}
\end{aligned}
\end{equation*}
(normal displacements have phase shift).

In this case from \eqref{eq:cyl} we get the system of linear algebraic equations
\begin{equation*}
\left\{
\begin{aligned}
&
-\frac{1}{6 R^2} \bigl( 3 U_\phi R^4 k_z^2 - 3 U_\phi R^4 k_z^2 \nu + 6 U_\phi R^2 \kf^2 + h^2 U_\phi R^2 k_z^2 \\&\qquad{}
- h^2 U_\phi R^2 k_z^2 \nu + 2 h^2 U_\phi \kf^2 + 3 R^3 U_z \kf k_z + 3 R^3 U_z \kf k_z \nu - 6 U_n \kf R^2 - h^2 U_n \kf \\&\qquad{}
- h^2 U_n \kf^3 - h^2 U_n \kf R^2 k_z^2 - 6 B_\phi R^2 \bigr) = 0 \\&
-\frac{1}{2} \bigl(R U_\phi \kf k_z - R U_\phi \kf k_z \nu - 2 U_z R^2 k_z^2 - U_z \kf^2 + U_z \kf^2 \nu + 2  \nu R U_n k_z + 2 B_z \bigr) =0 \\&
-\frac{1}{12 R^2} \bigl( 12 U_\phi \kf R^2 + 2 h^2 U_\phi \kf + 2 h^2 U_\phi \kf^3 + 2 h^2 U_\phi \kf R^2 k_z^2 \\&\qquad{}
+ 12 \nu R^3 U_z k_z - 12 U_n R^2 - h^2 U_n - 2 h^2 U_n \kf^2 - 2 h^2 U_n \nu R^2 k_z^2 - h^2 U_n R^4 k_z^4 \\&\qquad{}
- h^2 U_n \kf^4 - 2 h^2 U_n R^2 \kf^2 k_z^2 - 12 B_n R^2 \bigr) = 0
\end{aligned}
\right.
\end{equation*}

In the case of normal load ($B_\phi = 0$, $B_z = 0$, $B_n = -\frac{p R^2 (1-\nu)}{2h\mu}
$) the solution for the displacement intensities $U_m,\ m=\phi,z,n$ is
\begin{equation}\label{eq:el_solution}
\begin{aligned}
U_\phi &= -\frac{6 R^2 \kf p}{h \mu A} \bigl(-6 R^2 \kf^2 + h^2 \kf^4 \nu - 12 R^4 \kz^2 - 2 h^2 R^2 \kz^2 + 6 R^4 \kz^2 \nu + 6 R^4 \kz^2 \nu^2 - 3 h^2 \kf^2 R^2 \kz^2 \\&\qquad{}
+ 6 \kf^2 R^2 \nu + h^2 \kf^2 \nu - 2 h^2 R^4 \kz^4 + h^2 \kf^2 R^2 \kz^2 \nu - h^2 \kf^2 - h^2 \kf^4
\bigr) \\
U_z &= -\frac{6 p \kz R^3}{h \mu A} \bigl( - 3 h^2 \kf^2 \nu - 6 R^4 \kz^2 \nu + 6 R^4 \kz^2 \nu^2 + 6 R^2 \kf^2 - 6 \kf^2 R^2 \nu - 2 h^2 R^2 \kz^2 \nu + 2 R^2 \kz^2 \nu^2 h^2 \\&\qquad{}
+ h^2 \kf^2 + h^2 \kf^4 + h^2 \kf^4 \nu + h^2 \kf^2 R^2 \kz^2 + h^2 \kf^2 R^2 \kz^2 \nu
\bigr) \\
U_n &=  -\frac{6 R^2 p}{h \mu A} \bigl( - 2 h^2 \kf^4 + 2 h^2 \kf^4 \nu + 6 \kz^4 R^6 \nu - 6 \kz^4 R^6 + 2 h^2 \nu R^4 \kz^4 - 5 h^2 \kf^2 R^2 \kz^2 + 12 \kf^2 R^4 \kz^2 \nu \\&\qquad{}
- 12 \kf^2 R^4 \kz^2 + 6 R^2 \kf^4 \nu - 6 \kf^4 R^2 - h^2 \nu^2 R^2 \kz^2 \kf^2 - 2 h^2 R^4 \kz^4 + 2 h^2 \kf^2 R^2 \kz^2 \nu
\bigr) \\
A &= 6 h^2 \kf^4 - 72 \kz^4 R^6 \nu^2 - 24 \kz^4 R^4 h^2 \nu^2 + \kz^2 h^4 \kf^2 - 2 \kz^2 h^4 \kf^4 + \kz^2 h^4 \kf^6 + 24 \kz^2 R^2 \kf^6 h^2 \\&\qquad{}
- \kz^2 h^4 \kf^2 \nu + 2 \kz^4 R^2 h^4 \kf^2 \nu - 2 \kz^4 R^2 h^4 \kf^4 \nu - 4 \kz^4 R^2 h^4 \kf^2 + 4 \kz^4 R^2 h^4 \kf^4 + 24 \kz^6 R^6 \kf^2 h^2 \\&\qquad{}
+ 5 \kz^6 R^4 h^4 \kf^2 + 4 \kz^6 R^4 h^4 \nu - \kz^6 R^4 h^4 \kf^2 \nu + 72 \kz^4 R^6 + 6 \kz^8 R^8 h^2 + 2 \kz^8 R^6 h^4 + 6 \kf^8 h^2 \\&\qquad{}
+ 2 \kz^4 R^2 h^4 + 12 h^2 \kf^2 R^2 \kz^2 \nu + 36 \kz^4 R^4 \kf^4 h^2 - 2 \kz^4 R^2 h^4 \kf^2 \nu^2 + 2 \kz^2 h^4 \kf^4 \nu - \kz^2 h^4 \kf^6 \nu \\&\qquad{}
+ 12 \kz^6 R^6 h^2 \nu + 30 h^2 R^4 \kz^4 + 24 h^2 \kf^2 R^2 \kz^2 - 12 h^2 \kf^6 \\&\qquad{}
- 12 h^2 R^2 \kf^4 \kz^2 \nu - 36 h^2 R^4 \kf^2 \kz^4 - 48 h^2 \kf^4 R^2 \kz^2
\end{aligned}
\end{equation}

\subsection{Comparison with earlier theories}\label{ch:rot_comp}

Equations of the shells of revolution, derived by us, differ from the results of other authors. Orders of equations in different variants are the same, and these equations are too complicated to draw conclusions about their qualitative resemblances and distinctions. That's why let's consider only cylindrical shell and some of its characteristics: those, that have been considered in the section \ref{sec:rot_buil_cyl}.

\subsubsection{Operator coefficients}

Operator coefficients of system of equations for cylindrical shell derived by different authors are given by $\L=\L^0+\gamma\L^1$. Major terms $\L^0$ are the same:

\begin{equation*}
  \begin{aligned}
    L^0_{11} &= \frac{1}{2}\left( \dz^2 + 2\df^2 - \nu\dz^2 \right) &
    L^0_{12} &= L^0_{21} = \frac{1}{2}(1+\nu)\dz\df \\
    L^0_{22} &= \frac{1}{2}\left( 2\dz^2 + \df^2 - \nu\df^2 \right) &
    L^0_{23} &= L^0_{32} = \nu\dz \\
    L^0_{33} &= 1 &
    L^0_{31} &= L^0_{13} = \df.
  \end{aligned}
\end{equation*}
Smaller terms $\L^1$ are:
\begin{subequations}
\begin{gather*}
\intertext{V.\,V.\,Eliseev \cite{Eliseev_1999_en}}
  \begin{aligned}
    L^1_{11} &= 2 \left( \dz^2+2\df^2-\nu\dz^2 \right) &
    L^1_{12} &= L^1_{21} = 0 \\
    L^1_{22} &= 0 &
    L^1_{23} &= L^1_{32} = 0 \\
    L^1_{33} &= 1+\lap\lap-2\df^2-2\nu\dz^2 &
    L^1_{31} &= L^1_{13} = 2 \left( 1-\lap \right)\df
  \end{aligned} \\
\intertext{V.\,Z.\,Vlasov \cite{Vlasov_1951_naca}}
  \begin{aligned}
    L^1_{11} &= 0 &
    L^1_{12} &= L^1_{21} = 0 \\
    L^1_{22} &= 0 &
    L^1_{23} &= L^1_{32} = \frac{1}{2}\left( \df^2\dz-2\dz^3-\nu\df^2\dz \right) \\
    L^1_{33} &= 1+\lap\lap+2\df^2 &
    L^1_{31} &= L^1_{13} = \frac{\nu-3}{2}\df\dz^2
  \end{aligned} \\
\intertext{A.\,L.\,Goldenveiser \cite{Goldenveizer_1976_en} \& V.\,L.\,Biderman \cite{Biderman_1977_en}}
  \begin{aligned}
    L^1_{11} &= 2\dz^2+\df^2-2\nu\dz^2 &
    L^1_{12} &= L^1_{21} = 0 \\
    L^1_{22} &= 0 &
    L^1_{23} &= L^1_{32} = 0 \\
    L^1_{33} &= \lap\lap &
    L^1_{31} &= L^1_{13} = - \left( 2\dz^2+\df^2-\nu\dz^2 \right) \df
  \end{aligned} \\
\intertext{Yu.\,N.\,Novichkov \cite{Vibrations_1978_en}}
  \begin{aligned}
    L^1_{11} &= 2\dz^2+\df^2-2\nu\dz^2 &
    L^1_{12} &= L^1_{21} = 0 \\
    L^1_{22} &= 0 &
    L^1_{23} &= L^1_{32} = 0 \\
    L^1_{33} &= \lap\lap &
    L^1_{31} &= L^1_{13} = - \left( 2\dz^2+\df^2-2\nu\dz^2 \right) \df
  \end{aligned} \\
\intertext{W.\,Fl\"ugge \cite{Flugge_1962_en}}
  \begin{aligned}
    L^1_{11} &= \frac{3}{2}(\nu-1)\dz^2 &
    L^1_{12} &= L^1_{21} = 0 \\
    L^1_{22} &= \frac{1}{2}(\nu-1)\df^2 &
    L^1_{23} &= L^1_{32} = -\frac{1}{2} \left( 2\dz^2-\df^2+\nu\df^2 \right) \dz \\
    L^1_{33} &= 1+\lap\lap+2\df^2 &
    L^1_{31} &= L^1_{13} = -\frac{3-\nu}{2}\df\dz^2
  \end{aligned}
\end{gather*}
\end{subequations}

It's hard to give preference to any of these formulas without regarding their grounds and peculiarities of their derivation. All variants of smaller terms have the same order of derivatives, similar level of complexity. All of them are symmetric. More significant differences will appear later.

\subsubsection{Roots of characteristic equations}

In the same way, as it was done in section \ref{sec:rot_buil_cyl} let's examine the roots of characteristic equations in cases $m=0$ and $m=1$\footnote{These characteristics are not necessarily derived in the referred books. We have derived these characteristics ourselves, using the equations, given in the books.}.

\begin{subequations}
\begin{gather*}
\intertext{V.\,V.\,Eliseev \cite{Eliseev_1999_en}}
  \begin{aligned}
  \intertext{\quad$m=0$}
  k_{1,2,3,4}^{(0)}&=0 \\
  k_{5,6,7,8}^{(0)}&=\pm\sqrt{\nu \pm \sqrt{\gamma^{-1}+1}\sqrt{\nu^2-1}} \to \\
    &\to \pm \left( \gamma^{-1/4} \sqrt{\pm\sqrt{\nu^2-1}}
    + \gamma^{1/4} \frac{\nu}{2\sqrt{\pm\sqrt{\nu^2-1}}} + \dots \right) \\
  \intertext{\quad$m=1$}
  k_{1,2,3,4}^{(1)}&=0; \\
  k_{5,6,7,8}^{(1)}&=\pm\sqrt{\frac{1}{1+4\gamma}
    \left( 2+5\gamma+\nu+3\nu\gamma \pm \sqrt{A} \right) } \to \\
    &\to \pm \left( \gamma^{-1/4} \sqrt{\pm\sqrt{\nu^2-1}}
    + \gamma^{1/4} \frac{2+\nu}{2\sqrt{\pm\sqrt{\nu^2-1}}} + \dots \right) \\
  A&=-\gamma^{-1}-5-4\gamma+9\gamma^2+4\nu+ 22\nu\gamma+30\nu\gamma^2 \\&\quad{}
    +\nu^2\gamma^{-1}+9\nu^2+26\nu^2\gamma+25\nu^2\gamma^2
  \end{aligned} \\
\intertext{V.\,Z.\,Vlasov \cite{Vlasov_1951_naca}}
  \begin{aligned}
  \intertext{\quad$m=0$}
  k_{1,2,3,4}^{(0)}&= 0 \\
  k_{5,6,7,8}^{(0)}&= \pm\gamma^{-1/4}\sqrt{\frac{\nu\sqrt{\gamma}
    \pm\sqrt{-1+\nu^2+\gamma^2}}{-1+\gamma}} \to \\
    &\to \pm\left( \gamma^{-1/4}\sqrt{\pm\sqrt{\nu^2-1}}
    - \gamma^{1/4}\frac{\nu}{2\sqrt{\pm\sqrt{\nu^2-1}}}+\dots \right)
  \intertext{\quad$m=1$}
  k_{1,2,3}^{(1)}&= 0
  \end{aligned} \\
\intertext{A.\,L.\,Goldenveiser \cite{Goldenveizer_1976_en} \& V.\,L.\,Biderman \cite{Biderman_1977_en}}
  \begin{aligned}
  \intertext{\quad$m=0$}
  k_{1,2,3,4}^{(0)}&= 0 \\
  k_{5,6,7,8}^{(0)}&= \pm\gamma^{-1/4}\sqrt{\pm \sqrt{\nu^2-1}} \\
  \intertext{\quad$m=1$}
  k_{1,2,3,4}^{(1)}&= 0 \\
  k_{5,6,7,8}^{(1)}&= \pm\gamma^{-1/4}\sqrt{\frac{1+\gamma}{1+4\gamma}
    \left( 2\sqrt{\gamma}\pm\sqrt{-1+\nu^2+4\gamma\nu^2} \right)} \to \\
    &\to \pm\left( \gamma^{-1/4}\sqrt{\pm\sqrt{\nu^2-1}}
    + \gamma^{1/4}\frac{1}{\sqrt{\pm\sqrt{\nu^2-1}}}+\dots \right)
  \end{aligned} \\
\intertext{Yu.\,N.\,Novichkov \cite{Vibrations_1978_en}}
  \begin{aligned}
  \intertext{\quad$m=0$}
  k_{1,2,3,4}^{(0)}&= 0 \\
  k_{5,6,7,8}^{(0)}&= \pm\gamma^{-1/4}\sqrt{\pm\sqrt{\nu^2-1}} \\
  \intertext{\quad$m=1$}
  k_{1,2,3}^{(1)}&= 0
  \end{aligned} \\
\intertext{W.\,Fl\"ugge \cite{Flugge_1962_en}}
  \begin{aligned}
  \intertext{\quad$m=0$}
  k_{1,2,3,4}^{(0)}&= 0 \\
  k_{5,6,7,8}^{(0)}&= \pm\gamma^{-1/4}
    \sqrt{\frac{\nu\sqrt{\gamma}\pm\sqrt{-1+\nu^2+\gamma^2}}{-1+\gamma}} \to \\
    &\to \pm\left( \gamma^{-1/4}\sqrt{\pm\sqrt{\nu^2-1}}
    -\gamma^{1/4}\frac{\nu}{2\sqrt{\pm\sqrt{\nu^2-1}}}+\dots \right) \\
  \intertext{\quad$m=1$}
  k_{1,2,3,4}^{(1)}&= 0 \\
  k_{5,6,7,8}^{(1)}&= \pm\frac{1}{2}
    \sqrt{\frac{1}{\gamma\left( -1-2\gamma+3\gamma^2 \right)}
    \left( -A\gamma \pm \sqrt{B} \right)} \to \\
    &\to \pm\left( \gamma^{-1/4}\sqrt{\pm\sqrt{\nu^2-1}}
    +\gamma^{1/4}\frac{2-\nu}{2\sqrt{\pm\sqrt{\nu^2-1}}}+\dots \right) \\
  A&=8+11\gamma+9\gamma^2-4\nu-15\nu\gamma-9\nu\gamma^2\\
  B&=\left( A\gamma \right)^2-4\gamma\left( -2-4\gamma+6\gamma^2
  \right) \\&\quad{}
    \times \left( -2-8\gamma-6\gamma^2+2\nu^2+6\gamma\nu^2+6\nu^2\gamma^2
      +2\nu^2\gamma^3 \right)
  \end{aligned}
\end{gather*}
\end{subequations}

Here we can notice that when $m=0$ all variants provide four zero roots for $k$. It means that corresponding complementary solution of homogeneous problem would be given by $U=C_0+C_1\zeta+C_2\zeta^2+C_3\zeta^3$. Such polynomial solution resembles the solution of the problem of the Bernulli-Euler bending beam. Really, in axisymmetric case equations of cylindrical shell \eqref{eq:cyl} are similar to the beam equations; each lengthwise fiber behaves like a beam.

\label{balka_oops}In the case $m=1$ there are significant differences in the solutions for $k$. Some variants provide beam-like solution, and some have only three zero roots. Beam solution in this case describes the deformation of the whole cylinder like a beam. Seems that such solution should exist. In some books tubes are frequently treated as rods \cite{Svetlitsky_1982_en}. Absence of such solution may be interpreted as substantial weakness of the system.

Even more visible differences will be seen during comparing the \emph{solutions} of the systems \eqref{eq:cyl} by different authors with each other and with the solution of elasticity problem for 3-dimensional tube.

\subsubsection{Solutions}

Just like in the paragraph \ref{sssec:solution} we have built the solutions for $U_n$ under normal sinusoidal load in each of considered models. We have also examined the membrane (momentless) model that can be easily derived if we put $\gamma = 0$ and built the solution for the 3-dimensional cylindrical tube, loaded by the equivalent load on
\begin{itemize}
	\item internal surface: $p_{in} = p R / R_{in} = p R / \left( R - \frac{h}{2} \right)$
	\item outer surface: $p_{out} = p R / R_{out} = p R / \left( R + \frac{h}{2} \right)$
	\item both surfaces: $p_{in} = p_{out} = p R / \left(R_{in}+R_{out}\right) = p/2$,
\end{itemize}
where $p_{in}$ and $p_{out}$ are loads on correspondingly internal and external surfaces of 3D-cylinder, considered to be analogous to the only surface load in the shell.
 
The solution of such 3-dimensional problem has the form $u(\phi,z,r) = U(r)\exp{i \kf \phi + i \kz z}$ for each component of the displacement vector. To compare the intensity of radial displacement $U_r(r)$ with the normal displacement from shell models $U_n$ we take the averaged displacement $U_r = \frac{1}{h}\int_{R-h/2}^{R+h/2}{U_r(r)\,dr}$.

\begin{figure}[htbp]
	\includegraphics*[width=12cm]{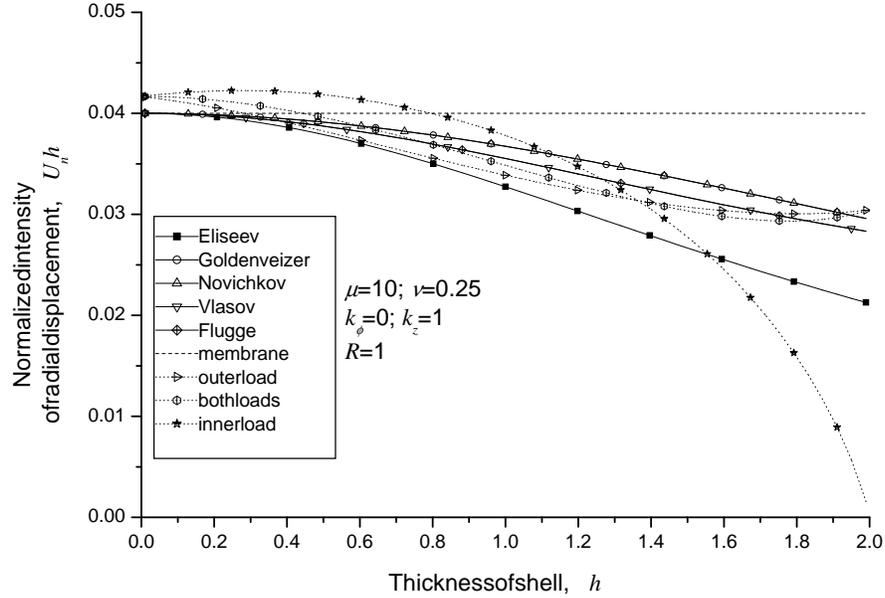}
	\caption{Eliseev's solution gives good estimate for displacements}\label{fig:n01}
\end{figure}

\begin{figure}[htbp]
	\includegraphics*[width=12cm]{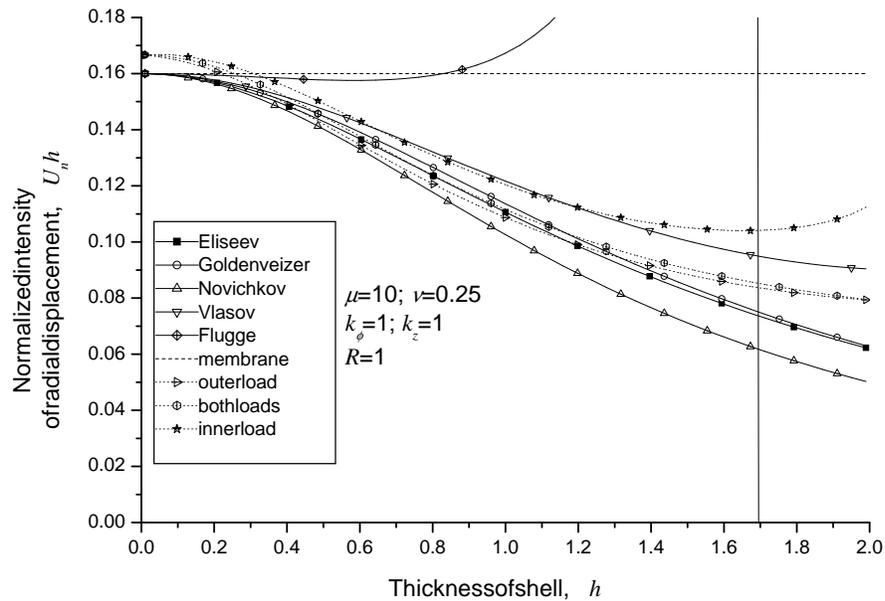}
	\caption{Inner load differs from the both other}\label{fig:n11}
\end{figure}

\begin{figure}[htbp]
	\includegraphics*[width=12cm]{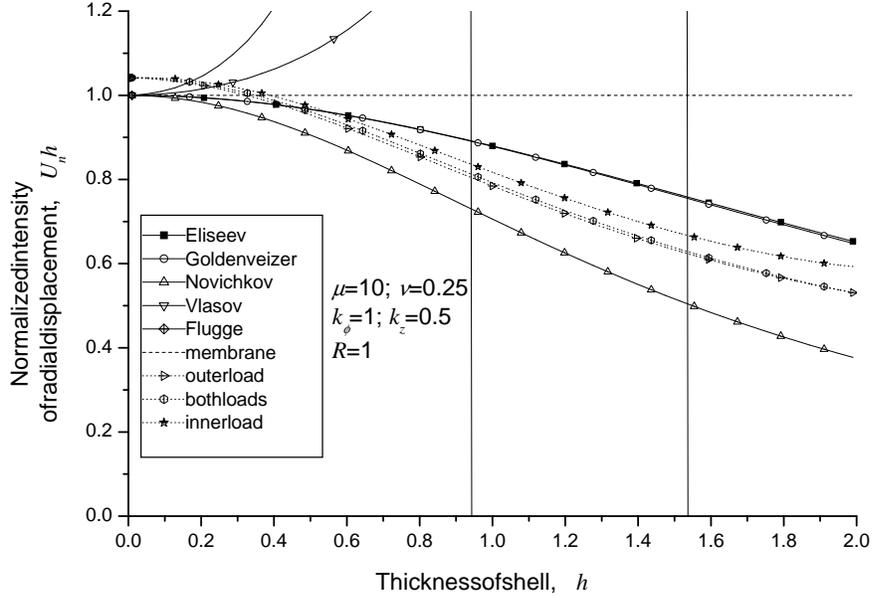}
	\caption{All but two shell models give very close solutions}\label{fig:n105}
\end{figure}

We have examined several cases of variability coefficients. They are shown on the figures  \ref{fig:n01}, \ref{fig:n11}, \ref{fig:n105}.


On these figures the dependence of the product of normal displacements and the thickness $U_n(h) h$ from thickness $h$ is shown. We choose $U_n(h) h$ instead of $U_n(h)$ because the function $U_n(h)$ is hyperbolic, has singularity in $h=0$ and spans across too big range in the ordinate axis, making the details of function behavior too small on the plot. Multiplying by $h$ removes hyperbolic aspect of the function and makes the plot more comprehensible.

So, first of all we may check that in all cases small thicknesses make the results of the shell and membrane models the same. That is natural, because when $\gamma \to 0$ all shell models lose their personal $\L^1$ terms and turn in the membrane model indeed.

But 3D-models give a bit higher values of radial displacement in some range of low $h$ --- rigidity occurs to be lower than in 2D-models. This may probably be explained by the less number of modeling suppositions, conditions, restraints on the nature of the deformation process in the classic 3D elasticity.

Next, we may see that membrane model always gives higher displacements than most of the shell models and 3D-models in higher range of $h$. That is caused by the higher rigidity of 3D and shell models, brought by bending (moment) terms.

In all cases we may also notice that 3D-cylinder with inner load differs from the both other 3D-models, especially in the higher range of $h$. We can conclude that the model with inner load seems to be somewhat special, because it leads to the singular load in limit case $h=2R$.

In the axissymmetric case ($\kf = 0$) we may see that Eliseev's solution stands apart from all other shell models, but it gives good estimate for displacements also in the very high $h$-range, reflecting the possibility of so special inner load case.

In the antisymmetric case ($\kf = 1$) all but two shell models give very close solutions. ``Outsiders'' are Vlasov's and Fl\"ugge's models --- they are discontinuous. Let's examine this discontinuity.

The expressions for the $U_n$ in both these models are rational functions as well as \eqref{eq:el_solution}. And due to the polynomial in $h$ in the denominator they may have discontinuities. In all models there is multiplier $h$ in the denominator of $U_n$, so this function is discontinuous in $h=0$. But there is also another point of discontinuity. Using MAPLE it is quite easy to find them. In the case $\kf=1,\ R=1,\ \nu=0.25$ these points are given by the equations:
\begin{equation*}
\begin{aligned}
	\text{Vlasov --- } &h_d: \quad (14 \kz^2 + 8 \kz^6 + 19 \kz^4 + 3)h_d^4 \\&\qquad\quad{}
		+ (144 - 96 \kz^6 - 96 \kz^2 - 336 \kz^4)h_d^2 - 1080 \kz^2 = 0 \\
	\text{Fl\"ugge --- } &h_d: \quad ( - 18 \kz^4 + 73 \kz^2 + 48 \kz^6) h_d^6 \\&\qquad\quad{}
		+ ( - 1212 \kz^2 - 1608 \kz^4 - 768 \kz^6 + 288) h_d^4 \\&\qquad\quad{}
			+ ( - 6912 + 8064 \kz^4 + 2304 \kz^6 - 4176 \kz^2) h_d^2 + 25920 \kz^2 = 0
\end{aligned}
\end{equation*}

\begin{figure}[htbp]
	\includegraphics*[width=12cm]{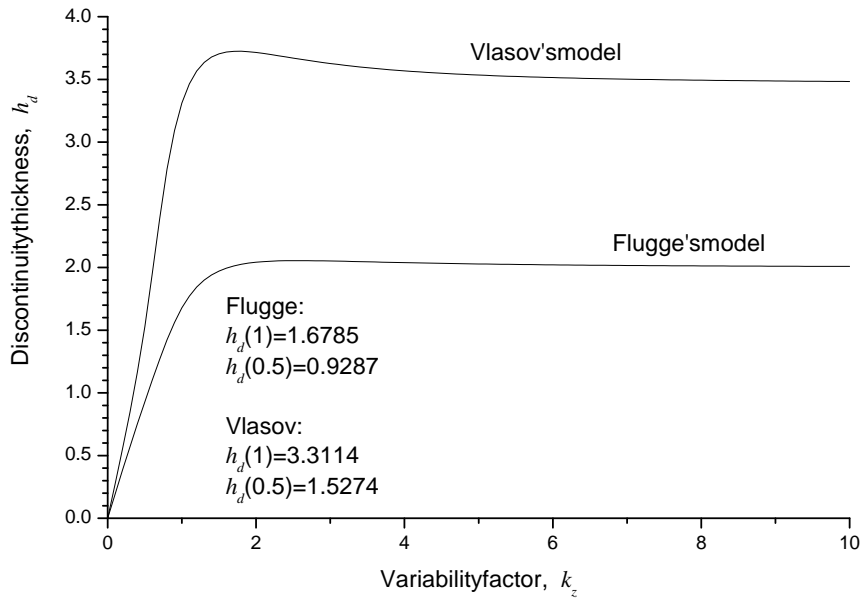}
	\caption{Discontinuity thickness dependent on $\kz$}\label{fig:discont_n}
\end{figure}

Though Vlasov's model gives biquadratic equation for $h_d$ with taken parameters, in general case there's no way to solve these equations analytically, therefore we may find the real roots numerically. On the figure \ref{fig:discont_n} there is the dependence of $h_d$, where the discontinuity takes place, from $\kz$. It should be noted also that the second elastic module $\mu$ has no influence on this relation.

We can see that discontinuity takes place in the possible range of $h \in (0; 2R)$ and corresponds with figures \ref{fig:n11}, \ref{fig:n105}.

%% file: concl.tex
\section{Conclusion}

In this study we have faced with high correlation of shell displacements with 3D ones also in thick shells, where any simplifying preconditioning numerical algorithms (see e.g. \cite{Xanthis_2000}) would not be useful, because of the acceptable condition number, and computational resources would be wasted without big improvement in precision.

On the other hand such approach to the shell theory doesn't give an idea about the connection of the stress tensors $\tta$ and $\tmu$ with 3D stress, therefore we cannot use these solutions within the well-developed strength conditions. This is the matter of the future studies as well as the derivation of other special cases of shell theory, e.g. shallow and narrow shells.

It is obvious that direct tensor algebra makes the analytical calculations much easier, bringing besides research advantages the pedagogical ones also.

The surprisingly inaccurate results, given by some of the considered shell models, make us understand that the shell theory is not well formed and closed even in its basis yet. Perhaps, even more unsunned surprises will be discovered during the foregoing future studies.